\documentclass[aps,prl,superscriptaddress,twocolumn]{revtex4}
\usepackage{graphicx}

\begin{document}



\title{Coherent Excitation of Rydberg Atoms in Thermal Vapor Microcells}


\author{H. K\"ubler}
\affiliation{5. Physikalisches Institut, Universit\"{a}t Stuttgart,
Pfaffenwaldring 57, 70550 Stuttgart}
\author{J. P. Shaffer}
\affiliation{5. Physikalisches Institut, Universit\"{a}t Stuttgart,
Pfaffenwaldring 57, 70550 Stuttgart}
\affiliation{Homer L. Dodge Department of Physics and Astronomy, The
University of Oklahoma, 440 W. Brooks St., Norman, OK 73019}

\author{T. Baluktsian}
\affiliation{5. Physikalisches Institut, Universit\"{a}t Stuttgart,
Pfaffenwaldring 57, 70550 Stuttgart}
\author{R. L\"ow}
\affiliation{5. Physikalisches Institut, Universit\"{a}t Stuttgart,
Pfaffenwaldring 57, 70550 Stuttgart}
\author{T. Pfau}
\affiliation{5. Physikalisches Institut, Universit\"{a}t Stuttgart,
Pfaffenwaldring 57, 70550 Stuttgart}
\email[]{t.pfau@physik.uni-stuttgart.de}
\homepage[]{www.pi5.uni-stuttgart.de}

\date{\today}

\begin{abstract}
%
The coherent control of mesoscopic ensembles of atoms and Rydberg
atom blockade are the basis for proposed quantum devices such as
integrable gates and single photon sources. So far, experimental
progress has been limited to complex experimental setups that use
ultracold atoms. Here, we show that coherence times of $\sim
100\,\mathrm{ns}$ are achievable with coherent Rydberg atom
spectroscopy in $\mu$m sized thermal vapor cells. We investigated
states with principle quantum numbers between 30 and 50. Our results
demonstrate that microcells with a size on the order of the blockade
radius, $\sim 2\,\mathrm{\mu m}$, at temperatures of
$100-300\,\mathrm{^\circ C}$ are robust, promising candidates to
investigate low dimensional strongly interacting Rydberg gases,
construct quantum gates and build single photon sources.
%
\end{abstract}


\maketitle


Recently, the mutual interaction between highly excited Rydberg
atoms in dense frozen samples has lead to the observation of Rydberg
atom excitation blockade \cite{Heidemann,Tong,Singer}. In Rydberg atom
blockade, the excitation of more than one Rydberg atom within a
blockade volume is suppressed as the mutual interaction between
Rydberg atoms at internuclear separations on the order of
micrometers shifts the atomic state out of resonance with a narrow
band excitation laser. The corresponding blockade radius,
$a_{\mathrm{block}}$, is on the order of several $\mu$m for Rydberg
states in the range of n$ = 30-50$. For example, the 32S state of Rb
excited by a 1 MHz bandwidth laser has $a_{\mathrm{block}} = 2
\,\mathrm{\mu  m}$ for an ensemble of atoms that do not move on the
timescale of excitation \cite{Schwettmann}. As the huge interaction
between individual Rydberg atoms can lead to controlled entanglement
of atomic ensembles, Rydberg atom blockade is the basis for
several proposals to realize photonic quantum devices, like single
photon sources and quantum gates \cite{Molmer,Jaksch}. The first
promising experimental steps toward this goal using individual well
localized pairs of ultracold atoms have been reported
\cite{Saffman,Grangier}. Experiments on collective entanglement of
ensembles of ultracold atoms have also been performed
\cite{Heidemann}.

A technologically interesting alternative approach to ultracold
atoms would be to realize Rydberg atom quantum photonic devices in
thermal Rb vapor microcells. For this idea, we envision arrays of
small blockade sized vapor cells (cavities) etched in glass that can be
connected by optical wave guides in a monolithic structure. In such
a device, Rydberg atom quantum gates may be realized with mesoscopic
ensembles of thermal atoms. Some of the advantages of this approach
are the ability to exploit advances in microstructuring technology
\cite{Sharping,Schmidt,newsandviews}, the relative simplicity of maintaining and
regenerating the sample, the collective enhancement of the laser
matter dynamics, and the scalability. We also point out here that
$a_{\mathrm{block}}$ has a strong scaling with the atomic separation $R$,
proportional to
$R^6$ in the case of Van der Waals interactions. This means that its
value for atoms frozen in place only decreases by approximately $2.7$
for a thermal distribution of atoms at $T = 300\,\mathrm{^\circ C}$,
since $a_{\mathrm{block}}\propto \sqrt[6]{\mathrm{Doppler\,width}}$.

An important point for using mesoscopic ensembles of atoms confined
to a blockade volume is that they can be used to realize so called
''superatoms'', where collective ground and excited states mimic the
behavior of an individual atom with coherent collectively enhanced
dynamics \cite{Heidemann,Lukin}. The collective behavior allows for
fast manipulations of the qubits and provides immunity to noise and
processing errors \cite{Molmer}. The requirements for collective
quantum dynamics in a mesoscopic ensemble of Rydberg atoms are a
density of ground state atoms much larger than
$(a_{\mathrm{block}})^{-3}$ and a timescale for the coherent
dynamics that is short compared to the time the atoms need to move
one excitation wavelength. The first requirement guarantees a
substantial enhancement of the collective dynamics. The second
requirement is necessary to assure that the dynamics of the ensemble
are collective and coherent. Thermal Rb vapors spatially confined to
$\sim \mu$m length scales can provide the required densities at
temperatures around $100-300 \,\mathrm{^{\circ}C}$. Due to their
thermal velocities coherent collective dynamics are expected to be
limited to a timescale of a few nanoseconds. The timescale
requirement can be met in a thermal vapor cell by using nanosecond
pulsed bandwidth limited laser excitation \cite{Appliedoptics}
provided there are no further dephasing or decoherence mechanisms
active on that timescale or shorter \cite{newsandviews,AdamsGighertz}.
Under similar conditions, strongly interacting, low dimensional, Rydberg gases can
be investigated. In this case, low dimensional means that some or all of the
geometric dimensions of a structure are on the order of or less than 
$a_{\mathrm{block}}$.

The very quality that Rydberg atoms possess which is the fundamental
root of the dipole blockade effect, namely a large polarizability,
also leads to their strong interaction with nearby walls and
electric fields. While the Casimir Polder effect between Rydberg
atoms and a metallic surface has been investigated \cite{Hinds}, the
effect of dielectric walls on Rydberg atoms is virtually unknown.
The effect of dielectric surfaces on Rydberg atoms are expected to
be stronger than a metal surface because a dielectric surface can
become charged and the surface interaction can be enhanced by the
resonant interaction with surface polaritons
\cite{Hinds,Barton,WileySipe,Ducloy}. The effect of the microcell's
walls can be a serious impediment to implementing them to make
quantum photonic devices.

In this paper we study coherent Rydberg excitation of a thermal
vapor in a confined geometry, a two dimensional microcell, using
Rydberg state electromagnetically induced transparency (EIT)
\cite{Adams}. The technique probes the Rydberg state energy shifts and
line broadening. We observe an EIT signal at $\sim 1\,\mu$m
separation between two confining dielectric quartz walls for some
states in the range n=30-50, contrary to common expectations of very broad and
massivly shifted Rydberg spectra
\cite{Hinds}. In particular, for 32S we observe linewidths on the
order of $\sim 12-16 \,\mathrm{MHz}$. This fact indicates that
coherent collective dynamics in a thermal microcell are not
fundamentally limited by wall induced decoherence or dephasing
effects provided the correct Rydberg state is chosen. The
corresponding coherence times also demonstrate that Rb vapor
microcells can be good candidates for realizing scalable arrays of
quantum gates. This is a significant step toward realizing an
entirely new system that can be used for quantum photonic devices,
such as single photon sources, quantum memories, and gates all based
on the same technology.

\section{Electromagnetically Induced Transparency in Microcells}


In all experiments described below, 
a wedge shaped vapor cell with a Rb reservoir attached to it is used.
The slope of the wedged gap was
interferometrically found with a Michelson interferometer to be
$10.5\,\mathrm{\mu m\,mm^{-1}}$ and
also indicated that the flats were in contact at the narrow gapped
edge of the cell. To make sure that the flats are contacting each
other, Newton rings were observed in the thin region of the wedge.
The maximum gap thickness was 500$\,\mu$m.

\begin{figure}
\includegraphics[width=\columnwidth]{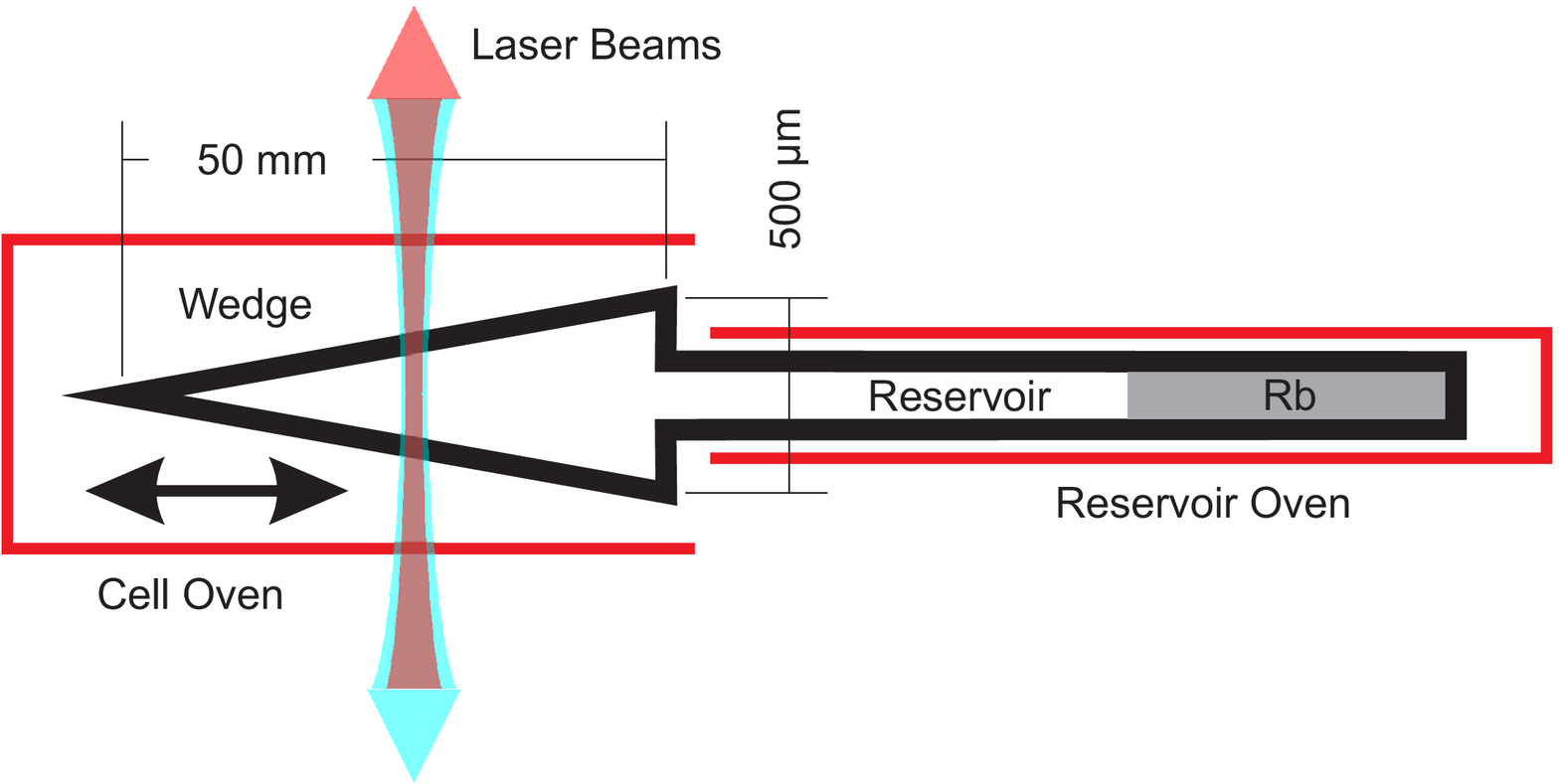}\newline
\includegraphics[width=\columnwidth]{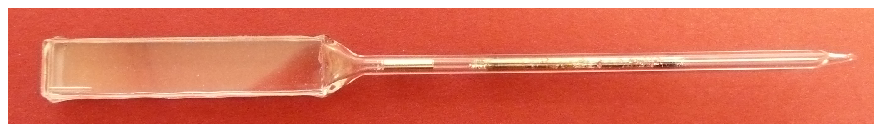}
\caption{Experimental setup. Two ovens allow for the control of the
temperature of the wedge and reservoir so that the Rb density can be
set. Moving the cell relative to the laser beams changes the
thickness of the Rb vapor layer. An image of the cell is
shown.\label{Fig:Setup}}
\end{figure}

The wedged part of the cell and the reservoir are placed in separate
ovens, Fig.~\ref{Fig:Setup}. This way the temperature of the wedge
and reservoir can be varied independently. The wedge is kept at
higher temperatures than the reservoir to prevent condensation of Rb
on its surface. By changing the temperatures of the two ovens, the
Rb vapor pressure and therefore the Rb density can be controlled.
The Rb vapor density used for the experiments was $1.6 \times
10^{14}\,$cm$^{-3}$. This corresponds to a Rb vapor pressure of
$4.4\,$mTorr in the microcell, which dominates all others.

EIT is chosen as the method to probe the behavior of the Rydberg
atoms in the cell because, as a coherent type of spectroscopy, it is
a direct measurement of the dephasing mechanisms of the Rydberg
state and of the Rydberg state energy shift. Since EIT depends on
quantum interference between the probe transition and probe plus
coupling laser transition, EIT is exquisitely sensitive to phase
disturbances on the coupling laser transition. Energy level shifts
of the Rydberg state cause the transmission window of the probe
laser to shift in frequency as well. To study the Rydberg states we
use EIT in a ladder configuration. We use a counterpropagating laser
geometry. The states involved in the EIT scheme used for this work
for a given Rydberg state, nl, are $5\mathrm{S_{1/2}}\rightarrow
5\mathrm{P_{3/2}}\rightarrow \mathrm{nl}$, Fig.~\ref{Fig:3lines}.
The first transition, called the probe transition, occurs at $\sim
780\,$nm while the Rydberg transition, called the coupling
transition, is at $\sim 480\,$nm. Principle quantum numbers between
$30$ and $50$ of $^{87}$Rb have been investigated both for $l=0$ and
$l=2$.

The peak Rabi frequency of the probe laser transitions is 5$\,$MHz
and the peak Rabi frequency of the coupling laser transition is
9$\,$MHz for 32S. The intensity of the coupling laser was adjusted
to maintain these Rabi frequencies for the other Rydberg states that
were studied. Signal detection in the narrow part of the cell
prevented us from decreasing the Rabi frequencies further. Under
these conditions, the total fraction of atoms in the Rydberg state
is $\sim 3 \times 10^{-5}$, yielding a Rydberg atom density of $\sim
5 \times 10^{9}\,$cm$^{-3}$ for 32S.
The density is chosen so that Rydberg-Rydberg interactions are negligible.
For 32S+32S, the expected Van der Waals interaction is less than
$\sim 1\,\mathrm{MHz}$.

The probe laser light that is transmitted through the cell is
coupled to a single mode fiber and measured by a photo multiplier
tube. The probe laser is locked to the
$5\mathrm{S_{1/2}}(F=2)\rightarrow 5\mathrm{P_{3/2}}(F=3)$
transition. The coupling laser light is intensity modulated using
and acousto-optic modulator (AOM) at a frequency of
$100\,\mathrm{kHz}$. The change in probe transmission correlated to
the presence of the coupling laser is measured with a lock-in
amplifier as a function of coupling laser frequency. The coupling
laser frequency is scanned at a rate of $\sim 400\,$MHz$\,$s$^{-1}$.
Up to 12,000 transmission curves are averaged per data point to
obtain the EIT signal. The probe and coupling laser frequencies are
referenced to an EIT signal that is obtained with a
$10\,\mathrm{cm}$ long Rb vapor cell in co- and counter propagating
laser geometry. We estimate the frequency stability of the EIT
signal at $\sim 2$MHz over the course of averaging $\sim$12,000
transmission curves.

\section{Results and Discussion}

\begin{figure}
\includegraphics[width=\columnwidth]{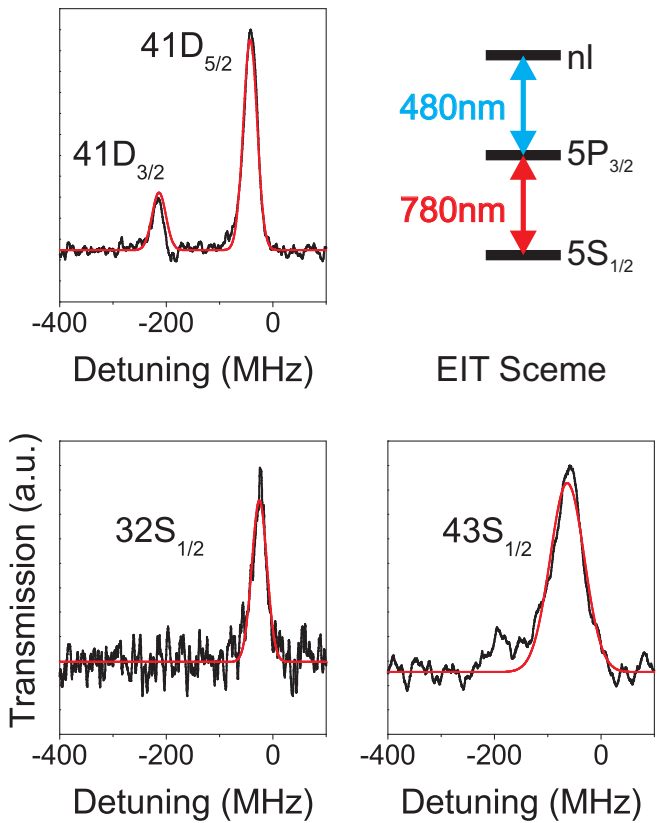}
\caption{EIT scheme and transmission signals as a function of
frequency. The 43S and 41D data were taken at wedge thicknesses of
$100\,\mu$m. The 32S data was taken at a wedge thickness of
$60\,\mu$m. The other experimental parameters are described in the
text. The lines are fits of the data to a Gaussian lineshape. The
EIT level scheme is also shown.\label{Fig:3lines}}
\end{figure}

First, we used a gap size large enough so that wall interactions were
negligible.
Fig.~\ref{Fig:3lines} shows the 43S and the 41D EIT signals in a
$100 \,\mathrm{\mu m}$ and the 32S EIT signal in a $60 \,
\mathrm{\mu m}$ thick section of the wedge. The fine structure of
the 41D state is clearly visible. For these states, wedge
thicknesses and experimental conditions already described, the
transmission signal has a linewidth of $\sim 12-16\,$MHz. The
linewidth in this part of the cell is determined by line broadening
due to ground state Rb atoms colliding with atoms in their Rydberg
states \cite{pressurebroadening}, laser frequency stability, transit
time broadening, and power broadening. In these regions of the cell,
the quartz walls have little effect on the EIT lineshape.

If we assume a thermal velocity of $v\approx 366 \,\mathrm{m\,
s^{-1}}$, consistent with the cell temperature of $195\mathrm{^\circ
C}$, the laser spot size leads to a transit time broadening of $\sim
4 \,$MHz. Collision broadening is dominated by collisions between
ground state atoms and Rydberg atoms. Collisions involving only
ground state and 5P$_{3/2}$ atoms have much smaller cross-sections
while collisions between 5P$_{3/2}$ atoms and Rydberg atoms are much
less frequent because the density of 5P$_{3/2}$ and Rydberg atoms is
much less than the ground state atoms. The collisional broadening
due to ground state atoms colliding with the Rydberg atoms has been
measured for Rb \cite{pressurebroadening}. For our temperatures and
densities, kept constant for these experiments, the collision
broadening is $\sim 11 \,\mathrm{MHz}$. Taking into account the Rabi
frequencies, laser stability over the time of the measurement,
transit time broadening and collisional broadening, we estimate a
minimum width of the EIT transmission signal of $\sim 14\,$MHz, in
excellent agreement with the measurements taken in parts of the
wedge where the wall interactions are negligible. Errors in the
measured linewidth are most likely due to differences in laser drift
and Rb gas density fluctuations during different data runs.


We have taken special care in determining the electric fields that
can lead to both line shifts and broadening. Charges
can buildup on the walls of the cell because it is a dielectric. The
electric field created by these charges can cause the Rydberg state
to shift in energy. If the distribution of charges is not uniform,
then the electric field will be inhomogeneous and will also cause a
broadening of the EIT signal. We made several measurements of the
electric field to verify that it did not dominate the Casmir-Polder
forces in the smaller sections of the wedge and that the
inhomogeneous electric fields can be overcome in order to
use microcells for quantum photonic devices. From these
measurements, we conservatively estimate the electric field at $<
4$V$\,$cm$^{-1}$ in the $\sim 1\,\mu$m part of the cell. The
electric field inhomogeneities were not detectable in the
experiments. The estimates of the electric field in the microcell
also suggest that field ionization of the Rydberg atom is not an issue
even in the $\sim 1 \,\mu$m parts of the wedge.
Magnetic field effects are not important in the work
here and can be safely neglected.

\begin{figure}

\includegraphics[width=\columnwidth]{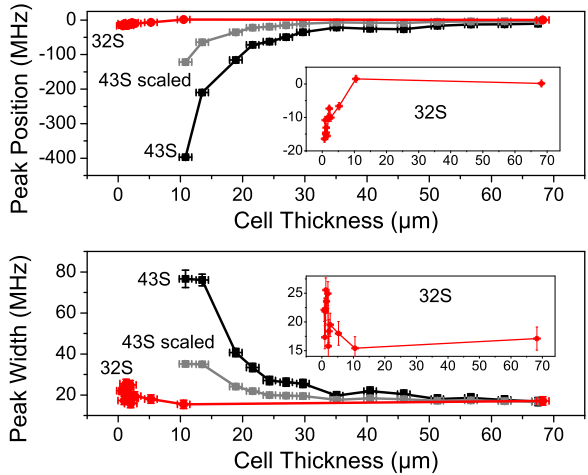}%

\caption{A comparison of the shift and broadening as a function of
wedge thickness for the 32S and 43S states. The curve labeled 43S
scaled is the data for the 43S state scaled to the dipole moment of
the 32S state. This curve shows that the weaker 32S interaction with
the cell wall is not due only to the differences in the dipole
moments. The scaling is described in more detail in the text. The
horizontal error bars show the uncertainty in the wedge thickness.
The vertical error bars show the fitting error and the point to
point variation as determined by variation in the measurements at
wedge thickness $> 50 \,\mu$m.\label{Fig:EIT43s}}
\end{figure}

In the thinner regions of the wedge, $< 20\,\mu$m, the wall
interactions dominate the dephasing and also shift the Rydberg state
energy. There are two components to the forces that the Rydberg
atoms experience due to the presence of the wall
\cite{Barton,WileySipe}. First, there is the interaction of the atom
with its image. This interaction essentially takes into account the
boundary conditions on the electromagnetic field due to the quartz
wall. It is $\sim 8\,$MHz for 43S and $\sim 2\,$MHz for 32S at R$=
2\,\mu$m \cite{Barton,Noordam}, where R is the distance between the
wall and atom. Secondly, the Rydberg atom can couple resonantly to
the surface polariton modes of the quartz. As this interaction
depends on the microscopic properties of the quartz, a detailed
estimate of its size is beyond the scope of this paper and will be
the subject of an additional work in the near future. The
interaction with surface polariton modes can be large for Rydberg
atoms because the energy intervals between dipole coupled Rydberg
states can match those of far infrared active polariton modes. It is
therefore important to chose a Rydberg state and microcell material
so that the resonant couplings between the Rydberg state and
polariton modes, including those from thermally excited polaritons,
are minimized in order to maximize the coherence time of the atoms
in a microcell. In our experiment, minimization of the overlap
between the resonances of the dielectric and the Rydberg atoms is
the essential point that allows for coherence times of roughly $100\,$ns
even at a cell thickness of $\sim 1 \,\mu$m.

The 43S state data exhibits strong effects from its interaction with
the quartz walls, as shown in Fig.~\ref{Fig:EIT43s}. At a cell temperature of
$195\,\mathrm{^\circ C}$ and a reservoir temperature of
$150\,\mathrm{^\circ C}$ we observed line broadenings up to $75
\,\mathrm{MHz}$ and a shift of $400 \, $MHz at a cell thickness of
$10\,\mathrm{\mu m}$. We attribute this behavior to a strong
interaction with surface polaritons. Note that some surface
polariton modes are thermally excited at $195\,\mathrm{^\circ C}$.
These modes can lead to other second order energy exchange pathways
between Rydberg states where one polariton can be annihilated and
another one can be created. These thermally activated pathways are
enabled by transitions with smaller differences in principle quantum
number and therefore possess larger dipole transition matrix
elements.

\begin{figure}
\includegraphics[width=\columnwidth]{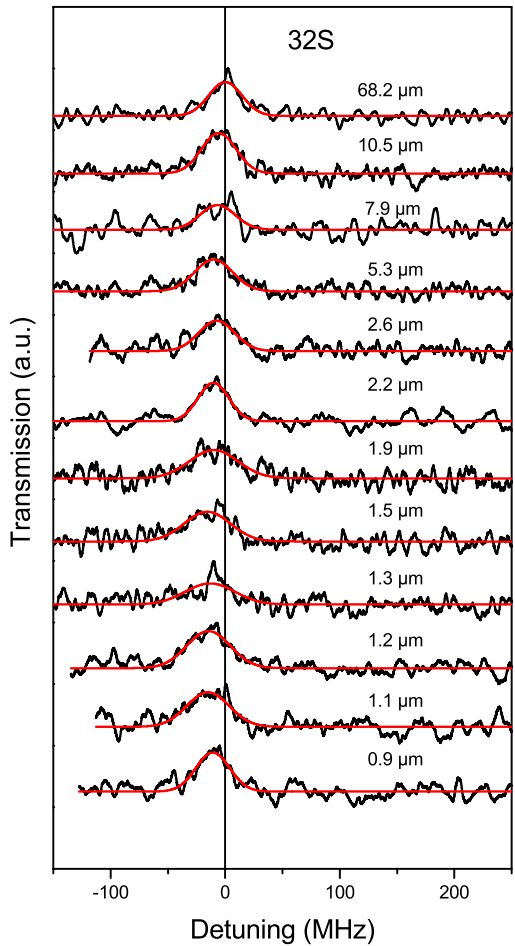}%
\caption{The transmission curves of the 32S state as a function of
frequency and thickness. The experimental parameters are described
in the text of the paper. The lines show Gaussian fits to the data.
The vertical line shows the zero of frequency as determined from the
reference signal obtained from the 10$\,$cm Rb cell. The uncertainty
in the wedge thickness is $\sim \pm 1 \,\mu$m. \label{Fig:EIT32s}}
\end{figure}

A spectrum as a function of wedge thickness for the 32S state is
shown in Fig.~\ref{Fig:EIT32s}. In contrast to the 43S data, we
observed narrow and only marginally shifted EIT lines all the way
down to $\sim 1.0\,\mathrm{\mu m}$, Fig.~\ref{Fig:EIT43s}. The
linewidths were $12-16\,$MHz and the maximum shift was $16
\,\mathrm{MHz}$ at a cell temperature of $195\mathrm{^\circ C}$. The
reservoir was kept at $150\,\mathrm{^\circ C}$. These conditions are
the same as for the 43S measurement. 32S has the smallest coupling
to the walls that we have observed so far. We attribute this result
to the fact that the 32S state has the fewest overlaps between its
allowed dipole transitions and infrared active transitions between
polariton states. The differences between the 32S and the 43S states
cannot be explained by the scaling of the respective dipole moments.
A plot of the 43S data scaled according to the n$^4$ dependence
\cite{Barton} of the interaction due to the different dipole moments
is shown in Fig.~\ref{Fig:EIT43s}. The magnitude of the shift of the
32S state is approximately what is expected from the interaction of
the atom with its image in the dielectric taking into account the
dominant transition to the nearest P states \cite{Barton,Noordam}.

\begin{figure}
\includegraphics[width=\columnwidth]{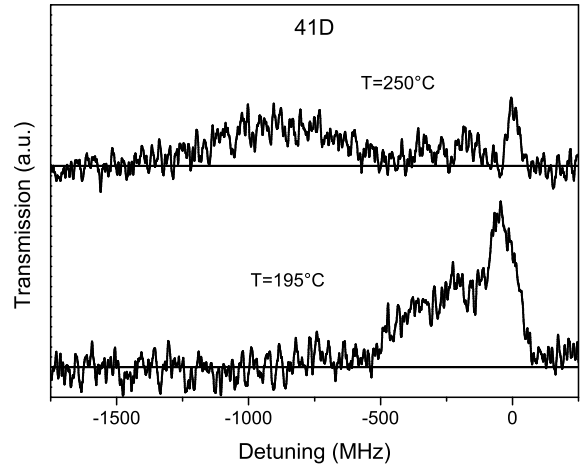}%
\caption{41D state. This plot shows the 41D state at two different
temperatures at a wedge thickness of 80$\,\mu$m.
\label{Fig:41Dstate}}
\end{figure}

The 41D state shows additional evidence for the presence of resonant
wall interactions, Fig.~\ref{Fig:41Dstate}. The interaction of the
atom with its image does not depend on polarization
\cite{Hinds,Barton,WileySipe}. The resonant interaction that
involves atoms interacting with the surface polaritons is
polarization dependent because of the dipole selection rules that
govern the allowed transitions for energy exchange with the wall.
For the 41D state we have observed a gap and temperature dependent
splitting of the m$_J$ sublevels of each of its fine structure
components, Fig.~\ref{Fig:41Dstate}. We observe one group of states
that is shifted and broadened similar to the 43S state, whereas
other states at the same time are not affected. This cannot be
explained by a Boltzmann factor weighting of the dominant dipole
transition. The temperature dependence of the dominant dipole
transition is very weak over the small range of temperatures we
investigated, $100-300\,\mathrm{^\circ C}$. The fact that different
m$_J$ states have different behavior supports our statement that the
resonant wall interactions are important since this has to result
from a polarization dependent interaction. Similar observations were
obtained for the 30D state.

Further systematic investigations of the details of the atom wall
coupling are clearly necessary and are the subject of further
studies. However, the fact that it is possible to observe Rydberg
atoms with linewidths and shifts on the $10\,\mathrm{MHz}$ level is
very encouraging for further investigations toward collective
coherent quantum dynamics in microcells. The data leads us to the
conclusion that the overlap between the polariton resonances of the
walls and the Rydberg atoms have to be reduced to obtain a coherence
time such that it is possible to realize Rydberg atom blockade in a
microcell.

\section{Conclusion}

To summarize, we have measured the dephasing and energy level shifts
experienced by Rydberg states in microcells with thicknesses on the
order of $a_{\mathrm{block}}$ using EIT. For 43S we observed a
significant wall induced line shift and broadening. The magnitude of
these effects in the $\sim \mathrm{\mu m}$ region of the cell
prevented us from obtaining a suitable signal for coherent
spectroscopy. For the 41D state we observed a group of states
strongly coupled to the surface and another one that remained narrow
- evidence for a dark state with respect to the atom wall coupling.
We obtained our best results for the 32S state. This state does not
have resonant overlap with the known polariton lines of quartz
\cite{Spitzer} and we observe an EIT signal that is suitable for
coherent manipulations of the dynamics in the $\sim 1\mathrm{\mu m}$
region of the cell. The frequency shifts and line broadening are
only on the order of a few $10 \,\mathrm{MHz}$. This is
significantly less than the bandwidth required for collective
coherent dynamics in a thermal vapor cell as discussed in the
introduction of this paper. Therefore we expect that collective
coherent dynamics are possible in a thermal microcell.

Our measurement is encouraging for further investigations that are
necessary for realizing single photon sources and other quantum
devices based on Rydberg atom blockade in thermal microcells. The
significant achievement of this study is the demonstration that
there are Rydberg states that can have long enough coherence times
to take advantage of the dipole blockade effect in thermal
microcells. In contrast to collective dynamics involving electronic
transitions, for example Rydberg states, the collective dynamics of
nuclear spin states are expected to have a significantly longer
coherence time, especially if spin protective coatings or buffer gas
loaded cells are
used. Future studies will utilize cells that make use of conductive
coatings to eliminate residual electric field effects and ones
coated and filled with buffer gas to enhance the coherence times of
nuclear spin states. Optimized cell materials and surface structures 
can also be used to
minimize the interactions of polaritons and Rydberg atoms.

\section{Methods}

The cell was fabricated from two quartz optical flats (Herasil 1) of thickness
of $5\,\mathrm{mm}$, cut to a size of $5\,\mathrm{cm}\times 1
\,\mathrm{cm}$. To form the cell, a spacer of $500 \,\mathrm{\mu m}$
thickness is placed between the flats to make a 5$\,$cm long wedge
with a maximum gap of 500$\,\mu$m. All the edges of the wedge,
except the one where the spacer is placed, are sealed by fusing the
glass. Next, the spacer is removed and a tube is connected to the
open side of the wedge and the wedge is sealed. The cell is pumped
through the tube to pressures below $10^{-6}\,\mathrm{mbar}$, the
tube is partially filled with Rb and is finally sealed off. The tube
acts as the reservoir.

The oven used to heat the wedge has 2 windows so that the light used
for the experiment can pass through the cell. This oven heats the
wedge uniformly without obstructing optical access. The Rb density
is monitored by absorption measurements on the Rb
$5\mathrm{S_{1/2}}\rightarrow 5\mathrm{P_{3/2}}$ transition. From
fits of the absorption spectra we were able to reproduce the wedge
temperature to within $\pm 5\,\mathrm{^\circ C}$. Therefore, we were
able to control the Rb density and optical density in the wedged
cell.

The coupling and probe beams pass through the oven windows, optical
flats and vapor in a counter propagating geometry, essentially
normal to the surface of the cell, Fig.~\ref{Fig:Setup}. The
position of the beams along the wedge can be changed by moving the
cell in and out of the oven heating the wedge on a translation
stage. This allows the thickness of the gap and consequently the Rb
vapor to be set for a given experiment. The accuracy of the cell
thickness is better than $1\,\mu$m.

The probe laser has a power of 16$\,\mu$W and is focussed to a beam
waist of $237\,\mathrm{\mu m}$. The coupling laser has a power of
800$\,\mu$W for 32S and is focussed to a beam waist of
$193\,\mathrm{\mu m}$. The laser-matter interaction region is
cylindrical with a diameter of 193$\,\mathrm{\mu m}$ and length
equal to the size of the wedge gap. The Rayleigh ranges of the
focussed beams are much longer than the cell gap width for all the
data presented here. The coupling laser has a bandwidth of 2$\,$MHz
over a ms timescale. The probe laser bandwidth is approximately the
same. The coupling laser bandwidth is more important because the
Rydberg linewidth is narrow compared to the laser bandwidth, thus
reducing the effective Rabi frequency delivered on the coupling
transition.

The reference cell is a $10\,\mathrm{cm}$ long Rb vapor cell in co-
and counter propagating laser geometry. To determine the reference
signal, a fraction of the probe laser is shifted with an AOM
200$\,$MHz to supply the probe light for the reference cell. The
coupling laser light used in the reference cell is derived from the
same source as that in the thin cell. This light is frequency
shifted 200$\,$MHz and amplitude modulated at 100$\,$kHz with a
$\pi$ phase shift relative to that used for experiments in the thin
cell. A second lock-in amplifier is used for this setup. The co- and
counter-propagating EIT signals provide a well defined calibration
of the frequency via the differential Doppler effect. Additionally,
each scan is triggered on the counter-propagating signal to provide
a stable locking point for the averaging of the scans.

The Stark shifts for alkali-atom Rydberg states are well known and
there are highly accurate methods to compute them \cite{Kleppner}.
Consequently, the field inside the cell can be estimated by
measuring the line splitting of a D state or the shift of an S
state. We observed a wall distance dependent line splitting for the
41D state from which we determined the electric field to be $< 3.6
\,\mathrm{V cm^{-1}}$ at $\sim 1 \,\mu$m wedge gap. The measured
electric field depended linearly on the wedge gap which is evidence
that the charges were on the surfaces of the wedge. This measurement
was done in the larger, $> 20 \, \mu$m, sections of the wedge to
avoid mixing the Stark shift caused by the surface charges with the
interaction between the dielectric wall and the Rydberg atoms. In
addition, for the 32S state, which was observed to have a weak
interaction with the walls, we observed only a small shift. This
data places an upper limit on the electric field in the $\sim 1
\,\mu$m part of the wedge at $\sim$4$\,$V$\,$cm$^{-1}$, consistent
with the electric field measurement performed using the 41D state
splittings. This estimate is an upper limit since it attributes all
the observed shift to an electric field. A third piece of data that
supports our field measurements is the fact that we do not observe
the blue shift of the m$_J = 1/2$ 41D$_{5/2}$ or the m$_J = 1/2$
41D$_{3/2}$ states. From the 41D Stark map this observation
indicates that the field is smaller than 1.2$\,$V$\,$cm$^{-1}$ at a
gap of $20\,\mu$m and suggests that the other measurements are upper
bounds and in fact the electric field may be smaller.

\section{Acknowledgments} We acknowledge fruitful discussions with
H.P. B\"uchler and H. Giessen as well as financial support from the
Landesstiftung Baden-W\"urttemberg. J. P. Shaffer acknowledges
support by the Alexander von Humboldt Foundation.

\section{Author contributions}
All authors contributed extensively to the work presented in this
paper.

\end{document}